\documentclass[11pt]{article} 
\usepackage{amssymb}
\usepackage{graphicx}

  \baselineskip \advance\textheight by \topskip
\large\normalsize
\newcommand{\be}{\begin{equation}}
\newcommand{\ee}{\end{equation}}
\newcommand{\ba}{\begin{array}}
\newcommand{\ea}{\end{array}}

\begin{document}
 \centerline{hep-th/0111107 \hfill CERN--TH/2001--314}

 \vspace{10mm} 
 \centerline{\LARGE\bf Holography of asymmetrically warped space-times}

 \medskip\bigskip
 \centerline{\large\bf Paolo Creminelli\footnote{E-mail address: {\tt Paolo.Creminelli@sns.it}}}
 \centerline{\em Theoretical Physics Division, CERN, CH-1211 Geneva 23, Switzerland,} 
 \centerline{\em Scuola Normale Superiore and INFN, Sezione di Pisa, Italy.}
 \vspace{5mm}
 \centerline{\large\bf Abstract}
 \begin{quote}\indent 
 We study the holographic dual of asymmetrically warped space-times, which are asymptotically AdS. The 
 self-tuning of the cosmological constant is reinterpreted as a cancellation of the visible sector stress-energy
 tensor by the contribution of a hidden CFT, charged under a spontaneously broken global symmetry. The apparent violation 
 of 4D causality due to bulk geodesics is
 justified by considering that the CFT feels the background metric as smeared out over a length of the order of the AdS radius.
 \end{quote}

\section{Introduction}
In the bulk of new extra-dimensional ideas, some attention has recently been devoted to asymmetrically warped
space-times \cite{Csaki:2001dm}: generalization of AdS solutions without 4D Lorentz symmetry, in which only the 
3D rotational invariance is maintained. There are two main reasons for this interest: first of all the violation of
Lorentz invariance appears through a variation of the speed of light moving in the bulk. This could be observable
studying the speed of gravitational waves or of any particle allowed to move in the extra dimension, {\em e.g.} 
right-handed neutrinos \cite{DeGouvea:2001mz}. The other reason is that these solutions can give
a hint towards the solution of the cosmological constant problem: these models have no low energy Lorentz
invariant description, so that it is conceivable to evade Weinberg's no-go theorem \cite{Weinberg:1989cp} on
the adjustment of the cosmological constant. In fact, explicit examples can be shown in which the only
maximally symmetric solutions describe a Minkowski metric on the brane.

In this note we want to study these backgrounds, which are asymptotically AdS (\footnote{Other solutions, which
are not asymptotically AdS, have been studied (see for example \cite{Grojean:2001pv}); we stick to the simplest 
cases in which the holographic dual is known. Lorentz non-invariance in asymmetrically warped spaces have been studied
also in the context of higher derivative gravity \cite{Nojiri:2001ae}.}), through the AdS/CFT correspondence.
In particular we want to understand how the two main features of interest of these spaces are reinterpreted in the dual 
theory, which is purely 4-dimensional. At first one may be puzzled about this point of view: how can we select
a 4D solution without cosmological constant? How can we have, in 4D, signals travelling {\em faster} than light? 
Do we loose 4D causality?
We will see that the answers to these questions are pretty simple, but we hope they can help to better understand, 
as in other cases \cite{Arkani-Hamed:2000ds,Rattazzi:2000hs}, the
phenomenology of these models and the limitations of the AdS/CFT correspondence. 
In Section \ref{Lambda} we study the simple example of a 5D charged black hole and give the holographic description
of the self-tuning of the cosmological constant. In Section \ref{causal} we study the signal transmission through
bulk geodesics and its dual interpretation.

\section{\label{Lambda} Holographic description of the adjustment of the cosmological constant}
Assuming that the 3D rotational invariance is preserved, the general solution of the Einstein equations in 
the presence of a negative cosmological constant $\Lambda_{\rm bulk} = -24 M^3/L^2$ ($M$ is the 5D Planck mass) 
and of a U(1) gauge field is given by the AdS--Reissner--Nordstr\"om metric:
\be
\label{eq:AdSSRN}
ds^2 = -h(r) dt^2 + h(r)^{-1} dr^2 +  \frac{r^2}{L^2} d\Sigma^2_k 
\ee
with
\be
\label{eq:AdSSRN2}
h(r) = k + \frac{r^2}{L^2} - \frac{\mu}{r^2} + \frac{Q^2}{r^4} \;.
\ee
As we are interested in 3D flat solutions we will take $k = 0$: $d\Sigma^2$ describes flat 3D sections. The metric 
depends only on the mass $\mu$ and the charge $Q$ of the black hole as implied by Birkhoff's theorem; $L^{-1}$ is the 
asymptotic AdS curvature.  If seen from the outside (large-$r$ region) the black hole singularity is shielded behind a horizon 
located at $r = r_h$, where $r_h$ is the largest root of the equation $h(r) = 0$. 

Now we introduce a brane in the region $r > r_h$: the brane joins two copies of the metric 
(\ref{eq:AdSSRN}) linked by a $\mathbb{Z}_2$ orbifold symmetry. The gauge field component $A_r$ is chosen to be 
$\mathbb{Z}_2$-even, while $A_\mu$ is odd. Notice that with this choice we can take the brane as neutral under 
the gauge symmetry. 
The motion of the brane in this background can
be studied using Israel's junction conditions to obtain the evolution of the induced metric \cite{Kraus:1999it}, 
which is of the FRW type
\be
\label{eq:FRW}
ds^2_{\rm brane} = -d\tau^2 + R^2(\tau) d\Sigma^2 \;.
\ee 
In particular we can look for static solutions so that the induced metric on the brane is the pure Minkowski space. We 
obtain \cite{Csaki:2001dm}:
\begin{itemize} 
\item If the pressure to energy ratio of the brane $\omega \equiv p/\rho$ satisfies $\omega \le - 1$, then we can find
a static solution in a given interval of $\rho$, choosing the mass and charge of the black hole.
\item The other 4D maximally symmetric solutions, namely 4D de Sitter and anti-de Sitter, are obtained only for 
$\mu = Q =0$ and $\omega = -1$.
\end{itemize}
In this way we see that the 4D Minkowski metric is the only maximally symmetric solution obtained for a 
continuous set of parameters, so that we can imagine an adjustment mechanism which changes the black hole
parameters to keep the effective 4D cosmological constant vanishing. 
If we look at the dynamical solutions we see that the black hole gives two contributions to the effective 
4D Einstein equations. One is a positive energy density proportional to $\mu$ and red-shifting as $R^{-4}$; the other 
is a negative energy density proportional to $Q^2$ and scaling as $R^{-6}$. 

It is interesting to find the holographic interpretation of the described solutions to understand the possible
mechanism of cancellation of the cosmological constant in a purely 4D language. It is known that a slice of AdS
terminated by a ``Planck brane'' is dual to a 4D theory with dynamical gravity, in which the matter living on
the brane is gravitationally coupled to a CFT with a large number of colors $N \sim (M L)^{3/2}$ \cite{maldacena}.
The ``unusual'' choice of taking $A_r$ $\mathbb{Z}_2$-even 
makes $A_\mu$ vanishing on the brane. In this way the U(1) symmetry is not gauged in the 4D picture, as it would be 
for the other parity assignment \cite{Arkani-Hamed:2000ds}, but it 
remains global, though spontaneously broken by the brane itself. In fact we can check that the 5D equations of motion
give a massless normalizable zero mode
for the 4D scalar $A_r$, with a constant profile in the $5^{\rm th}$ dimension: we interpret it as the Goldstone boson 
$\pi$ of the broken symmetry. The theory is invariant, as it must, under a constant shift $c$ of the Goldstone field 
\be
\label{eq:gauge}
\pi(x^\mu) \rightarrow \pi(x^\mu) + c \;,
\ee
because this corresponds to a pure gauge $A_r$= constant in the 5D picture.
Notice that, the global symmetry being broken, we have not the problem to reconcile the 4D gravitational violation of global symmetries
({\it e.g.} global charge disappearance into a black hole) with the corresponding 5D local invariance, which we 
expect to be respected by gravity \cite{Silverstein:2001kk}.

Now we can reinterpret the two contributions described above to the 4D Einstein equations:
\be
\label{eq:rhoCFT}
\rho_{\rm CFT} + \rho_{\rm GB} \propto \frac\mu{R^4} - \frac{Q^2}{R^6} \;.
\ee

The first one is the usual scale-invariant ($\rho \propto T^4$) contribution of a hot CFT ($p = \rho/3$) \cite{Gubser:2001vj}.
The second one comes from the interaction of the hot CFT with the Goldstone boson. The relevant part of the 
Lagrangian, neglecting higher dimension operators with two or more $\pi$'s, will be of the form
\be
\label{eq:lagrGold}
{\cal{L}}_{\rm GB} = \frac12 (\partial_\mu\pi)^2 + F_\pi^{-1} J_{\rm CFT}^\mu \partial_\mu\pi \;,
\ee
where $J_{\rm CFT}$ is the CFT current associated with the broken symmetry and $F_\pi$ is the Goldstone particle decay constant.
Notice that this parameter has the dimension of a mass: this will lead to a non-scale-invariant contribution to the stress-energy 
tensor. 
The total conserved current is given by $F_{\pi} \partial\pi + J$.
The gravitational solution has a black hole of charge $Q$ in the bulk and a mirror black
hole of charge $-Q$ in the other $\mathbb{Z}_2$ symmetric half space (the charge is $\mathbb{Z}_2$-odd): this describes 
a hot CFT, which is charged under the global symmetry. Notice however that the total conserved charge is zero: if we start from
a configuration with no gauge field (neutral black holes), the emission of charged particles from
the Planck brane leads to the solution with charged black holes and an electric field $F_{tr}$ on the brane; 
as 4D charge is conserved, also this final state must have vanishing total charge. This means that the CFT 
charge density $\rho_Q$ is compensated by the Goldstone boson contribution: 
\be
\label{eq:charge}
F_\pi \partial_t \pi + \rho_Q = 0  \;. 
\ee
The global charge neutrality can also be seen as a consequence of Gauss law in the bulk, which relates the bulk charge 
({\em i.e.} the CFT charge) with the integral on the brane of the field strength component $F_{tr}$ (which gives the
Goldstone boson contribution $\propto \partial_t \pi$). It is easy to check that this leads to eq.~(\ref{eq:charge}). 

If we now calculate the stress-energy tensor from the Lagrangian (\ref{eq:lagrGold}) and use eq.~(\ref{eq:charge}), we get 
a contribution
\be
\rho = p = - \frac12 \frac{\rho_Q^2}{F_\pi^2} \;.
\ee 
Having $p = \rho$ this energy density red-shifts as $R^{-6}$; this can also be understood by noting that $\rho_{\rm GB}$
is proportional to the square of the charge density, which goes as $R^{-3}$. The 4D description is compatible with the 5D
gravity equations and gives the holographic interpretation of the stiff matter contribution found in
\cite{Biswas:2001sh}.

 
Summarizing, the 4D description is the following: the matter on the brane describes a visible sector in which the global
symmetry is broken; bulk physics describes a hidden sector formed by a hot CFT which interacts with the visible one
only through gravity and higher dimension operators. Some charge is transferred from the visible to the dark sector and
this charge acts as a source for the Goldstone boson, so that it becomes time-dependent.

To have a stationary flat solution we must require that the total energy and pressure for the three components described 
above cancel out. The parameters (energy density; pressure) are respectively of the form:
\begin{eqnarray}
{\rm Brane\;matter} & : & \rho_{\rm brane} \cdot (1; \widetilde\omega) \\
{\rm CFT} & : & \rho_{\rm CFT} \cdot (1 ; 1/3) \\
{\rm Goldstone\;boson} & : & -|\rho_{\rm GB}| \cdot (1 ; 1) \;.
\end{eqnarray}
For brane matter we mean all that is added to the fine-tuned positive tension brane with 
$\Lambda_{\rm brane} = 24 M^3/L$: its contribution is cancelled by the bulk curvature and does not enter in the 
holographic dual theory. We see that varying the two parameters that describe the CFT we can compensate 
the energy density and pressure on the brane\footnote{If we take $Q=0$ we have only one free parameter and every 
static solution must be fine-tuned \cite{Csaki:2001dm}.}. This can be done with some restrictions: it is easy to check 
in the 5D picture that the sum $\rho_{\rm CFT} + \rho_{\rm Gold.}$ is always positive and that the hidden sector
contribution satisfies the weak energy condition $\rho + p \ge 0$ (\footnote{This actually holds only if the matter on
the brane gives a subdominant contribution with respect to the brane tension; this is anyway required if we want
to neglect higher derivatives of the Ricci tensor in the 4D dual.}). Therefore the visible sector matter must have 
negative energy density and $\rho + p \le 0$: $\widetilde\omega \ge -1$. The brane tension
does not contribute to $\rho + p$, so that  also the total energy-momentum tensor of the brane (brane tension + visible matter) 
in the 5D picture violates the
weak energy condition. This gives the bound $\omega \le -1$ for the whole brane discussed before \cite{Csaki:2001dm}.

It is easy to see that the violation of the weak energy condition \cite{Cline:2001yt} is pretty generic because in 
the 4D language there is a cancellation of $\rho + p$ among the various components. From this 4D point of view it is obvious
that de Sitter or anti-de Sitter are obtainable only with a precise choice of the parameters: namely we have to 
``turn off'' the CFT setting $\mu = Q =0 $ and put pure tension on the brane. Only in this way does the 
total stress-energy tensor have $p = -\rho$ as required by ${\rm dS_4}$ or ${\rm AdS_4}$ solutions.    
    
In the holographic counterpart of the 5D model the possibility to evade Weinberg's no-go theorem is seen in
a slightly different way. Now the bulk dynamics is reinterpreted in a purely 4D language so that the theory {\em is} 
Lorentz invariant; nevertheless the static configurations {\em are not} Lorentz invariant: the CFT is, in particular, 
in a thermal state. 

This point is quite general: the bulk metrics that violate the $SO(3,1)$ symmetry are holographically mapped into
4D states that are not Lorentz invariant. We notice that this is rather particular, because from the 4D point
of view it is obvious that this Lorentz violating effects disappear for sufficiently high energy: locally
the theory {\em is} Lorentz invariant as implied by the equivalence principle. As stressed in \cite{Chung:2000ji}, this
is not what happens in general when we look at the 4D effective action obtained as reduction of a higher dimensional 
theory: in this case the theory may never behave as Lorentz invariant, even in the UV limit. The point is 
that we are considering rather particular spaces that are asymptotically AdS$_5$ so that it is obvious that they 
behave as Lorentz invariant in the high energy limit, which is sensitive only to the asymptotic region.

In the same reference \cite{Chung:2000ji} a cosmological problem for extra-dimensional theories was considered: as standard
4D cosmology must address the question of why the Universe is very close to the spatial flatness, now
we have the further task to justify the approximate $SO(3,1)$ symmetry of the bulk. In the particular models 
we are studying, these two problems are deeply linked and they can be solved in the same way\footnote{This relation was briefly 
discussed in \cite{Anchordoqui:2000du}.}. It is in fact clear,
from the 4D point of view, that a period of inflation solves not only the flatness problem (as usual) but
also the second one, because every possible non-invariant initial state is red-shifted away. From
the 5D picture we look at the inflationary phase as a movement of the Planck brane towards the UV region which,
as we said, is asymptotically AdS$_5$ and therefore $SO(3,1)$-invariant.

Even if we have studied only the simple example given by the metric (\ref{eq:AdSSRN}), we expect all our discussion to remain 
qualitatively true for other asymmetrically warped space-times, asymptotic to AdS. In particular the self-tuning 
of the cosmological constant will always turn into a cancellation among the various 4D energy components.

\section{\label{causal} Causality on the brane and in the bulk}
A characteristic property of asymmetrically warped space-times (see metric (\ref{eq:AdSSRN}) for example) is that
the speed of light is different at different points along the extra dimension. The
holographic interpretation of this fact is simple: signals propagating in the bulk are viewed as excitations of the 
CFT and there is no reason to expect these excitations to have the speed of light. The problem is to understand 
what the interpretation of signals moving {\em faster than light} is: they violate causality from a 4D perspective. 
How is this possible?

First of all we have to distinguish two different cases. In the presence of a naked singularity it is possible that the
speed of light on the surfaces at fixed $r$ 
\be
\label{eq:speed}
c(r) = \frac{L}r h(r)^{1/2}
\ee
is growing, moving far from the brane ({\em i.e.} going to smaller $r$). This
happens in the AdS--Reissner--Nordstr\"om metric if the charge is high enough: there are no positive root of the equation
$h(r) = 0$ and the singularity is naked. In
this case, even if the brane is fixed at a given value of $r$, with flat induced metric, bulk signals can travel faster 
than those remaining on the brane. The different speed of light and gravitational signals could give experimental signatures 
in the forecoming gravitational wave experiments: electromagnetic and gravitational waves emitted by a supernova 
would arrive at different times \cite{Csaki:2001dm}. The presence of a naked singularity is however
problematic from both the gravitational and the CFT side: it is difficult to discuss if and how the gravitational singularity
will be resolved\footnote{A similar problem appeared in other extra-dimensional attempts to cancel the 4D cosmological 
constant \cite{Arkani-Hamed:2000eg}.}, while from the 4D point of view it is likely that we are trying to describe a theory 
which is not in its true vacuum so that causality could be violated. In the following we will study the cases in 
which the singularities are shielded by horizons, so that the gravitational and CFT pictures are under control. 
If the singularities are shielded, the speed of light is decreasing going to smaller $r$ and a geodesic escaping into the
bulk cannot have a turning point and come back to the starting value of $r$. Nevertheless we can always have geodesics 
that connect two points on the brane protruding into the bulk if the brane moves (or more generally if it is bent) into
the bulk itself. In this case the difference in signal propagation is deeply linked with the
induced metric on the brane so that experimental signals can only come from cosmology.  


To face the problem we study an example that is simpler than the metrics previously discussed, but equivalent
with respect to this aspect of apparent causality violation. We take the pure AdS$_5$ metric and a positive 
tension brane with matter living on it, which moves in the 5D space obeying, as usual, Israel's junction conditions. 
From the brane point of view we have an induced metric of the form (\ref{eq:FRW}), with the only difference that 
now the 4D Einstein equations have no additional piece due to the bulk CFT. Let us study the propagation of bulk 
signals in this simple setting. 
Using the AdS$_5$ parametrization 
\be
\label{eq:AdSmetric}
ds^2 = \frac{L^2}{z^2} (dx^\mu dx_\mu + dz^2) \;,
\ee
the distance $r$ travelled by a gravitational signal between two points $A$ and $B$ is obviously given 
by\footnote{Notice that there is no turning point: the geodesic always moves in a fixed $r$ direction. However, the 
brane motion into the bulk may be such that the gravitational signal leaves the brane and reintersects it at a
subsequent time.}
\be
\label{eq:geod}
(z_B - z_A)^2 + r^2 = (t_B - t_A)^2  \;.
\ee 
In these coordinates the induced metric on a brane moving in the AdS background is of the form 
\be
\label{eq:induced}
ds^2_{\rm brane} = -d\tau^2 + \frac{L^2}{z_{\rm brane}^2 (\tau)} dx^i dx_i  \;:
\ee
the FRW scale factor $R$ is given by $L/z_{\rm brane}$ and   
\be
\label{eq:proper}
dt^2 = \frac{1}{R^2} (1 + L^2 H^2) d\tau^2  \;,
\ee
where $H$ is the Hubble parameter.
Let us take the two points $A$ and $B$ lying on the brane: from (\ref{eq:geod}) and (\ref{eq:proper}) we get
\cite{Caldwell:2001ja}
\be
\label{eq:gravhor}
r_g = \left(\left[\int_A^B \frac{d\tau}{R} \sqrt{1+ L^2 H^2} \right]^2 - 
\left[\int_A^B \frac{d\tau}{R} L H \right]^2\right)^{1/2}  \;.
\ee
This defines a ``gravitational horizon'': the maximum distance that can be covered from $t_A$ to $t_B$ through
signals moving in the bulk. We notice that this distance is always greater than the standard one, 
obtained by considering signals travelling on the brane (let's say ``light'' signals):
\be
\label{eq:lighthor}
r_\gamma = \int^B_A \frac{d \tau}{R} \;.
\ee
Notice that in obtaining eq.~(\ref{eq:gravhor}) we must assume that the brane motion is such that 
the geodesics can connect points $A$ and $B$ remaining inside the physical space, {\em i.e.} 
$z \geq z_{\rm brane}$. It easy to show that this is true if $dH/d\tau \leq 0$, which translates into the weak
energy condition $\rho + p \geq 0$ \cite{Ishihara:2001nf}. The inflating Universe $\rho + p =0$ is the 
borderline situation: bulk and brane geodesics coincide as implied by the fact that eqs (\ref{eq:gravhor}) 
and (\ref{eq:lighthor}) give the same result.

The problem is that the 4D holographic counterpart lives on the metric (\ref{eq:induced}) so that it defines the causal 
horizon by (\ref{eq:lighthor}), while we have seen that in the AdS picture the causal horizon is greater and given by 
(\ref{eq:gravhor}). 
We expect the two equivalent theories to have the same causal horizon, because this fixes the possibility
for a space-time event to affect another one. How can we reconcile the two points of view?

The solution is evident if one goes back to the definition of the duality. The introduction of the Planck brane
can be seen as a way of integrating out all degrees of freedom of the CFT above the scale $1/L$: the 4D theory is 
defined with a Wilsonian cutoff given by the inverse of the AdS radius. In
this way the conformal theory feels the gravitational background as smeared out over a typical length scale $L$ 
and this justifies the apparent discrepancy between (\ref{eq:gravhor}) and (\ref{eq:lighthor}).
For example eq.~(\ref{eq:gravhor}) can be specified for a standard power law evolution of the form 
$R(\tau) = R_0 \tau^\alpha$ with $\alpha < 1$ to give, for $H \ll L^{-1}$
\be
\label{eq:goalpha}
\frac{r_g}{r_\gamma} \simeq  1 + (L H_B)^2 \frac{1-\alpha}{2 (1+\alpha)} 
\left(\frac{R_B}{R_A}\right)^{1+\frac{1}{\alpha}}.
\ee
A significant result can be obtained on the CFT side if we consider that its excitations feel a scale factor $R(\tau)$
averaged over a time $L$. Always in the limit $H \ll L^{-1}$ we have
\be
\label{eq:average}
\bar{R}(\tau) \simeq R(\tau) \left(1  + (L H)^2 \frac{\alpha-1}{24 \alpha}\right)  \;:  
\ee 
calculating the causal horizon (\ref{eq:lighthor}) with $\bar{R}(\tau)$ instead of $R(\tau)$ 
the result (\ref{eq:goalpha}) is obtained up to a factor of order one. Actually it is not possible to establish a precise
correspondence because the details of the smearing procedure are not known, but we clearly see that there is no
obvious contradiction between the two ways of looking at the same theory. The same remains true if we consider
more complicated 5D metric as (\ref{eq:AdSSRN}): now the propagation of bulk signals is described by excitations
of the hot CFT, but the apparent violations of causality can be explained in the same way.

Notice that a consistent difference between the two horizons is obtained only if the typical curvature of the
induced metric ${\cal{R}}$ is greater than the AdS curvature. This regime is beyond the possibility of the CFT
description because an infinite number of operators involving higher derivatives of the metric should be taken 
into account. This is another indication that the theory is coarse-grained over the length scale $L$.

\subsection{The full AdS case}
How much of what of we said remains true in the pure AdS case, in which no brane is present and the 4D gravity is non-dynamical?
In general to understand if the causality structure given on a fixed surface by the induced metric is the same
as that given by the bulk metric, we must check if the surface has extrinsic curvature. In particular the
two structures do not coincide if \cite{Ishihara:2001nf} 
\be
\label{eq:extrK}
K_{ab} k^a k^b \neq 0  \;,
\ee 
where $K_{ab}$ is the extrinsic curvature of the surface and $k$ is a null vector for the induced metric. In this
case the null geodesics do not lie on the surface so that the causal structures are different.  

In the pure AdS case the bulk metric can be cast into the form \cite{Henningson:1998gx} 
\be
\label{eq:bulkmetric}
G_{ij} dx^i dx^j = \frac{L^2}{z^2} \left(g_{\mu\nu}(x^\mu,z) dx^\mu dx^\nu + dz^2\right)  \;,
\ee  
where $g$ tends to a given metric $g_{(0)}$ at the boundary ($z = 0$), which belongs to the conformal structure on which
the CFT lives. We can obtain an expansion for $g$, using the bulk Einstein equations, of the form\footnote{For our
discussion the presence of terms in $\log z$ \cite{Henningson:1998gx} is irrelevant.} 
\be
\label{eq:metricexp}
g = g_{(0)} + z^2 g_{(2)} + z^4 g_{(4)} \ldots \;,
\ee
with $g_{(n)}$ an expression with $n$ derivatives of the metric $g_{(0)}$. 

The extrinsic curvature of a surface at fixed small $z$ can be calculated to give at leading order
\be
\label{eq:extrinsic}
K_{\mu\nu} = - \frac{L}{z^2} g_{(0) \mu\nu} + {\cal{O}}(z^2).
\ee
As a function of the induced metric $\gamma$ on the surface at fixed $z$ we get
\be
\label{eq:extrinsic2}
K_{\mu\nu} \simeq  - \frac1L \gamma_{\mu\nu} + L \cdot g_{(2) \mu\nu}  \;.  
\ee
The first term is not important because, being proportional to the induced metric, it vanishes if applied to
null vectors; the second one is given by
\be
\label{eq:g2}
g_{(2)\mu\nu} = \frac1{2} \left(R_{\mu\nu} -\frac{1}{6} R \cdot g_{(0)\mu\nu}\right) \;,
\ee
where $R_{\mu\nu}$ and $R$ are the Ricci tensor and the scalar curvature of $g_{(0)}$.
We obtain for the component $K^{\rm NULL}$ of the extrinsic curvature non-vanishing on the null vectors
\be
\label{eq:extint}
K^{\rm NULL}_{\mu\nu} = \frac{L}2 R_{\mu\nu} + {\cal{O}}(z^2)\;.
\ee
The Ricci tensor for the metric $g_{(0)}$ is the same, at leading order in $z$, as that for the induced 
metric $\gamma$ so that we are relating the extrinsic and intrinsic curvatures of a surface at fixed $z$. The expression 
(\ref{eq:extint}) is valid for any metric $g_{(0)}$ we choose in the same conformal 
class, for any (sufficiently small) $z$ and therefore for the boundary $z=0$. We see that for any surface at fixed $z$
geodesics starting on the surface do not remain on it: the causal structure is different from that implied
by the induced metric $\gamma$.
The meaning of this remains the same as in the presence of a brane: whatever surface at fixed $z$ we choose, the CFT 
living on it, conjectured to be dual to the gravity theory inside \cite{Balasubramanian:1999jd}, does not know the 
details of the metric: it is always blurred over a length scale of order $L$. 
This is another way to look at the   
holographic idea that the boundary theory has only a limited storage capability: one degree of freedom per Planck 
volume \cite{Susskind:1998dq}. 
In fact a theory with $N^2$ fields living on a lattice of spacing $L$ has a number of degrees of freedom in a 3D 
volume $V$
\be
\label{eq:dof}
N_{\rm dof} \sim N^2 \frac{V}{L^3} \sim \frac{V}{G_5} \;.
\ee 
To conclude we stress that even if eq.~(\ref{eq:extint}) is valid for any $z$ and implies the finite number of 
degrees of freedom per Planck volume, in the limit $z \rightarrow 0$ the corrections to the causal structure due 
to bulk geodesics vanish. This follows from the fact that the scalar curvature of the surfaces tends to zero,
making the advantage of the shortcuts smaller and smaller. 
In this way, after removing the cut-off, the CFT feels exactly the conformal structure $g_{(0)}$ belongs to.
 
\section{Conclusions}
We have studied the holographic dual of asymmetrically warped space-times that are asymptotically AdS. The 
self-tuning of the cosmological constant is reinterpreted in a much more conventional (and unappealing!) way: the stress-energy tensor 
of the matter living on the brane is cancelled by a ``dark sector'' formed by a strongly coupled CFT, which is charged under
a spontaneously broken global symmetry. 
The possibility to detect a difference in light and gravitational waves speed is possible only in the presence of a naked singularity,
which is likely to describe a CFT in a wrong vacuum. In the presence of horizon-shielded singularities the difference between 
the causal 
horizon defined using only geodesics on the brane and that defined in the full space is justified
in the dual theory as a smearing of the 4D metric over a length scale of the order of the AdS radius. This difference,
however, is always small when the 4D dual makes sense. 

\paragraph{Acknowledgements} I thank Riccardo Rattazzi for many useful comments and Alessandro Strumia for constructive criticisms.
This work is partially supported by EC under TMR contract HPRN-CT-2000-00148.

\end{document}